\documentclass[twocolumn,prl,superscriptaddress]{revtex4-2}

\usepackage{amsmath,amssymb}
\usepackage{graphicx}
\usepackage{float}
\usepackage[noend]{algpseudocode}
\newcounter{algorithm}
\renewcommand{\thealgorithm}{\arabic{algorithm}}
\usepackage{xcolor}
\usepackage{subcaption}
\usepackage{hyperref}

\begin{document}

\title{Efficient Classical Simulation of Heuristic Peaked Quantum Circuits}

\author{David Kremer}
\email{david.kremer@ibm.com}
\affiliation{IBM Quantum, IBM T.J.Watson Research Center, Yorktown Heights, NY 10598}

\author{Nicolas Dupuis}
\affiliation{IBM Quantum, IBM T.J.Watson Research Center, Yorktown Heights, NY 10598}

\date{\today}

\begin{abstract}
Peaked quantum circuits, whose output distribution is sharply concentrated on a single bitstring, have emerged as a promising candidate for verifiable quantum advantage, as the correctness of the quantum output can be checked by simply comparing against the known peak. Recent work by Gharibyan et al.\ \href{https://arxiv.org/abs/2510.25838}{[arXiv:2510.25838]} claimed heuristic quantum advantage using peaked circuits executed on Quantinuum's 56-qubit H2 processor.
These peaked circuits concentrate their output on a single hidden bitstring by training a shallow simulable circuit variationally and inserting an obfuscated permutation to increase the depth to a level that makes classical simulation intractable, with estimated runtimes of years for the largest instances.
We show that these circuits can be efficiently simulated classically. We describe a method that efficiently performs a full tensor network contraction, allowing near-exact sampling and extraction of the peaked bitstring. The method exploits the mirrored structure of the circuit and iteratively cancels both halves into a Matrix Product Operator (MPO), and avoids the obfuscated permutation by greedily reducing the MPO bond dimension through a process we call ``unswapping.'' The method can fully contract and extract the peak of the largest circuit in approximately one hour on a single GPU, around half the time it took to run on the quantum hardware.
\end{abstract}

\maketitle

\section{Introduction}
\label{sec:introduction}

Demonstrating a computational advantage of quantum hardware over classical methods remains one of the central goals of quantum computing research. A key challenge is that any such demonstration requires not only that the quantum device outperforms the best known classical methods, but also that the quantum output can be rigorously verified. This has motivated calls for more rigorous frameworks for evaluating quantum advantage claims~\cite{lanes_framework}, and the creation of the Quantum Advantage Tracker~\cite{qa_tracker} to systematically collect and compare results.

Peaked circuits, first proposed by Aaronson and Zhang~\cite{aaronson_peaked}, offer an approach that addresses the verification problem directly. These circuits are constructed so that their output distribution is sharply concentrated on a single bitstring that has a measurement probability orders of magnitude higher than the rest. This bitstring is known at the time of circuit construction but is, in principle, computationally hard to extract from the circuit description without executing it on quantum hardware. This makes peaked circuits an appealing candidate for verifiable advantage: the answer obtained from quantum execution can be validated simply by checking against the known peak bitstring, with no expensive classical post-processing required.

In practice, however, the classical hardness of peaked circuits depends critically on how they are constructed. In the original proposal~\cite{aaronson_peaked}, the peak is induced by variationally adjusting circuit parameters, an approach that produces circuits with peaked distributions but does not scale beyond circuit sizes amenable to direct statevector simulation or standard tensor network methods. Tensor network methods, in particular Matrix Product States (MPS) and Matrix Product Operators (MPO)~\cite{vidal_2003, schollwock_review, orus_review}, are among the most powerful tools for classical simulation of quantum circuits, as they can efficiently represent states with limited entanglement by compressing the quantum state into a network of low-rank tensors. Their computational cost is controlled by the \emph{bond dimension}, which grows with the entanglement in the system. However, this cost can also be inflated by features of the circuit that do not correspond to genuine entanglement, such as permutations of qubit lines or long-range gates that scramble the ordering of the tensor network~\cite{zhou_limits_2020}. As we will show, this distinction between real entanglement and artificial bond dimension inflation is central to understanding why the peaked circuits of Ref.~\cite{bluequbit_peaked} are in fact classically tractable.

More recently, Gharibyan et al.~\cite{bluequbit_peaked} addressed the scalability limitation by combining the variational peaking technique with a mirror circuit construction and a suite of obfuscation transformations. The resulting circuits embed a $UU^\dagger$ identity block that dramatically increases depth, while tensor patch optimization, angle sweeping, masking, and swap insertions are applied to prevent classical methods from detecting and cancelling the identity. The authors benchmark several state-of-the-art classical simulation strategies, including MPS methods, tensor network belief propagation, and Pauli path simulation, and report that all fail to scale beyond roughly 700 two-qubit gates.

Among the classical strategies considered in Ref.~\cite{bluequbit_peaked}, the one most relevant to our work is what the authors call the ``Middle MPO Attack.'' This approach directly targets the mirrored structure of the circuit: it initializes a Matrix Product Operator (MPO) as the identity at the circuit midpoint and iteratively absorbs gates from both halves, exploiting the fact that the underlying $UU^\dagger$ structure should in principle allow the MPO to remain low-rank throughout. For unobfuscated mirror circuits, this is highly effective: the two halves cancel as they are absorbed, and the MPO never grows beyond manageable sizes. However, the authors show that the combination of swap transformations, angle sweeping, and masking is sufficient to defeat this strategy: the permutations introduced by the swap obfuscation cause the MPO bond dimensions to grow uncontrollably and prevent compression. The key insight of the present work is that this obstacle can be overcome.

Our method follows the same general principle as the Middle MPO Attack, splitting the circuit and contracting both halves into a central MPO, but introduces an additional step to handle the permutation structure that causes the naive approach to fail. Rather than attempting to compress the MPO directly after absorption, we apply a greedy procedure we call \emph{unswapping}, which identifies and extracts the hidden permutations from the MPO by iteratively testing swap operations on the bonds with the largest dimension. This decomposes the MPO as $M = P_L \tilde{M} P_R$, where $P_L$ and $P_R$ are permutations and $\tilde{M}$ is a reduced MPO with low permutation content. The extracted permutations are then propagated back into the remaining circuit through a rewiring step that restores alignment between the circuit and the compressed MPO, allowing the absorption to continue.

This iterative cycle of absorption, unswapping, and rewiring allows the full circuit to be contracted while keeping the MPO size manageable throughout. Once all layers have been absorbed, we apply the resulting MPO to the all-zero state to obtain a Matrix Product State (MPS), from which we sample directly. The method requires no approximation beyond a small singular value cutoff for numerical stability during compression, and thus produces near-exact samples from the ideal circuit.

We apply this method to the largest circuit presented in Ref.~\cite{bluequbit_peaked}, a 56 qubit circuit with 1,917 two-qubit gates, and recover the peak bitstring in approximately one hour on a single GPU (Nvidia A100, 80~GB), roughly half the time the circuit took to execute on the quantum hardware. This demonstrates that the obfuscation strategies employed in the peaked circuit construction are insufficient to prevent efficient classical simulation, and that the claimed quantum advantage does not hold for these instances.

\section{Method}
\label{sec:method1}

\begin{figure*}[t]
    \centering
    \includegraphics[width=\textwidth]{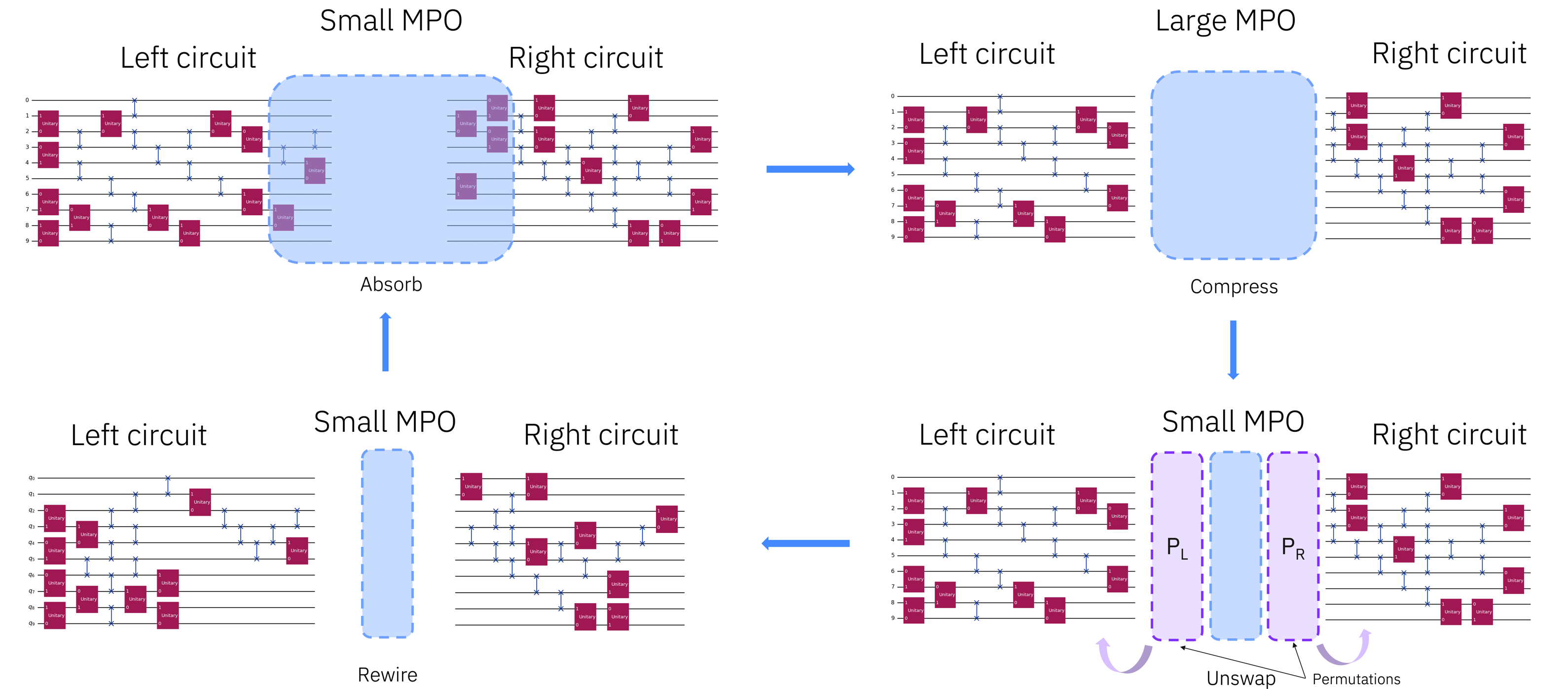}
    \caption{Overview of the iterative contraction procedure. Starting from a small MPO between the left and right circuits, the method cycles through three stages: (1)~absorption of circuit layers into the MPO, which causes it to grow; (2)~unswapping, which extracts permutations $P_L$ and $P_R$ and reduces the MPO to a smaller $\tilde{M}$; and (3)~rewiring, which propagates the extracted permutations into the remaining circuits and re-transpiles to linear connectivity.}
    \label{fig:overview}
\end{figure*}

We now describe the full contraction procedure. The method consists of three phases: circuit preparation, iterative contraction, and sampling. An overview of the full process is shown in Fig.~\ref{fig:overview}.

\subsection{Circuit Preparation}
\label{sec:preparation}

The peaked circuits in Ref.~\cite{bluequbit_peaked} are defined on an all-to-all connectivity native to the trapped-ion hardware. To enable efficient tensor network contraction, we first transpile the circuit to a linear (nearest-neighbor) connectivity by decomposing long-range gates into sequences of SWAP gates and nearest-neighbor operations. This transpilation introduces additional gates but ensures that the circuit can be represented and contracted as a one-dimensional tensor network.

\begin{figure*}[t]
    \centering
    \begin{subfigure}[t]{0.295\textwidth}
        \centering
        \includegraphics[width=\textwidth]{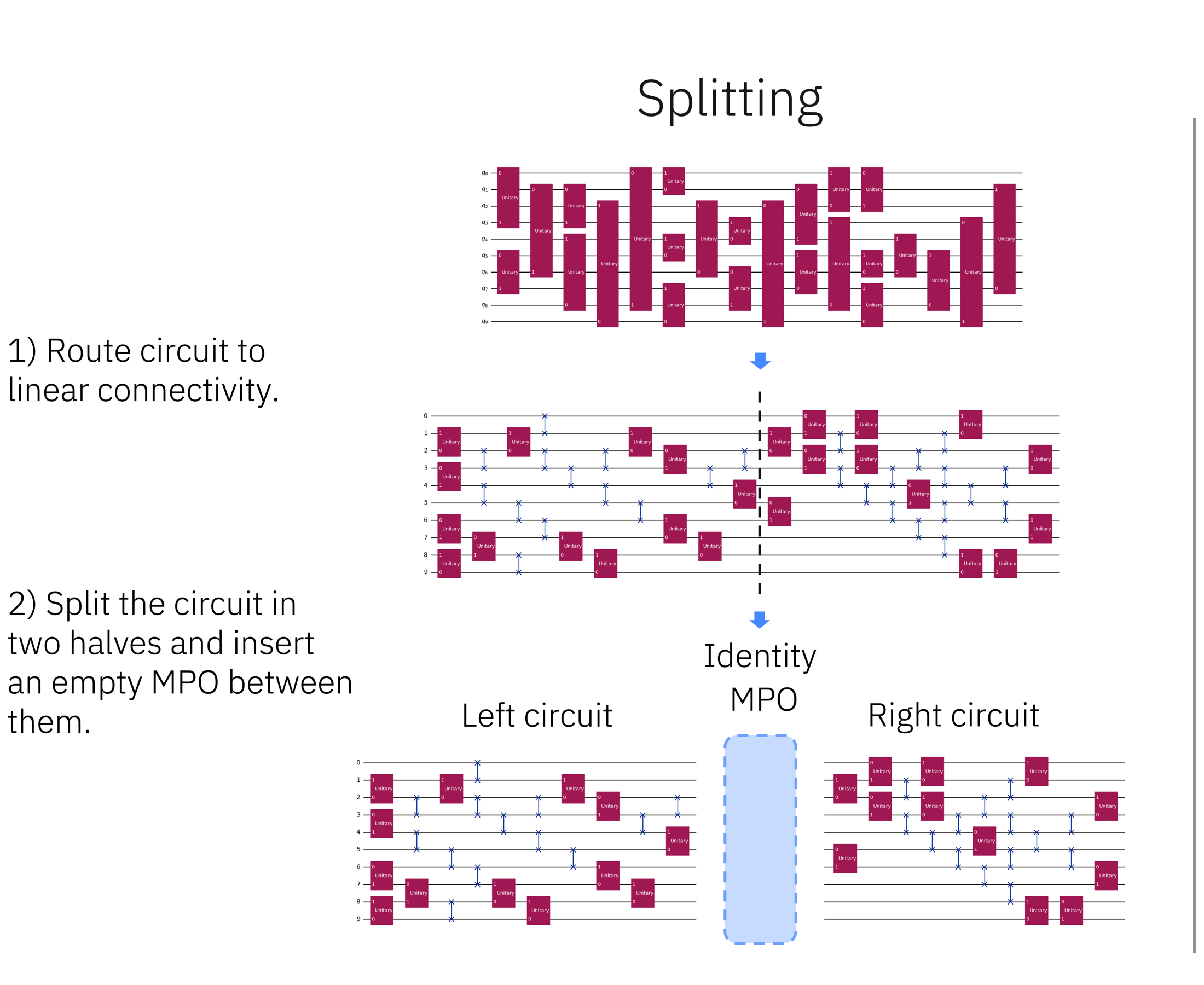}
        \caption{Splitting}
        \label{fig:splitting}
    \end{subfigure}
    \hfill
    \begin{subfigure}[t]{0.355\textwidth}
        \centering
        \includegraphics[width=\textwidth]{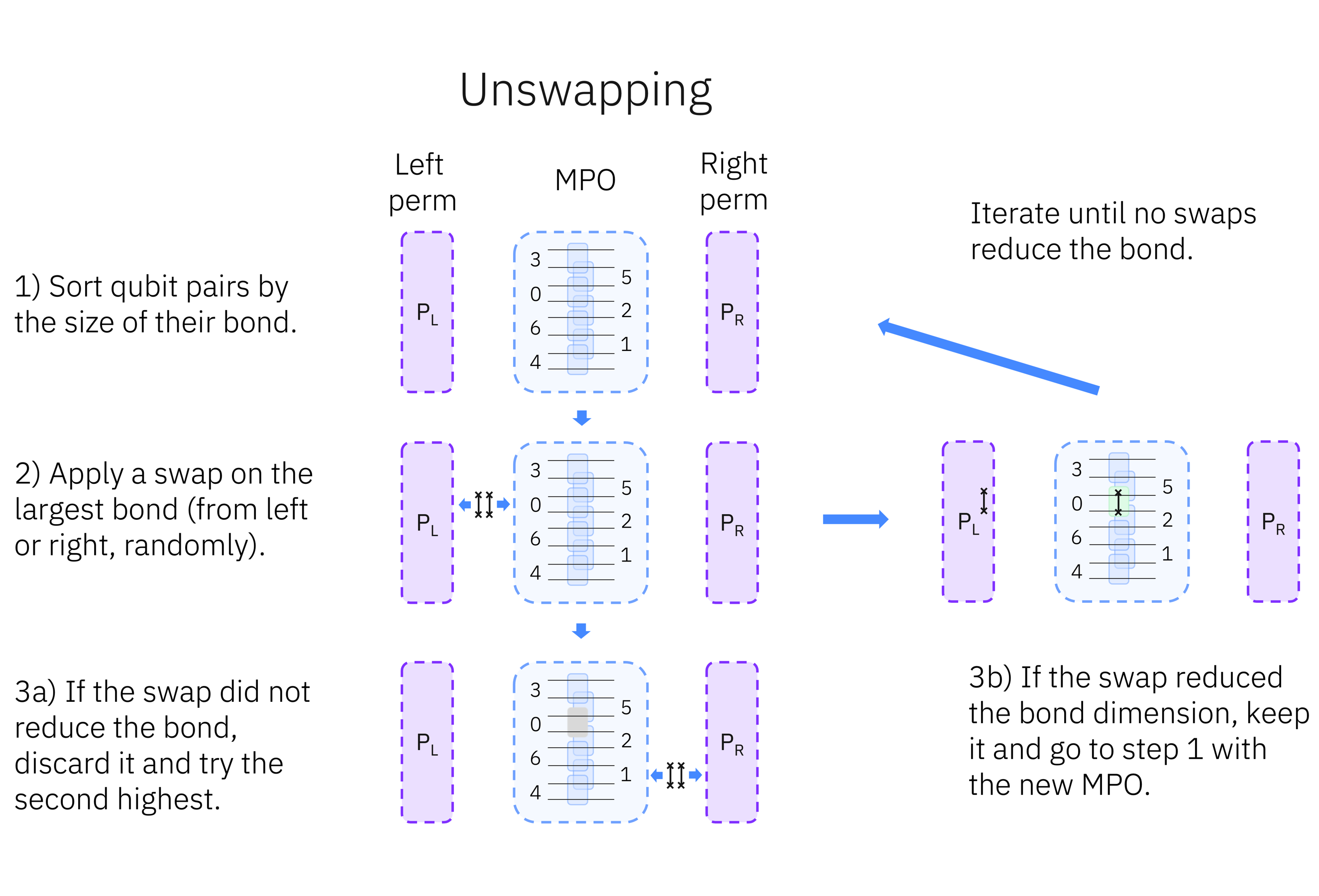}
        \caption{Unswapping}
        \label{fig:unswapping}
    \end{subfigure}
    \hfill
    \begin{subfigure}[t]{0.311\textwidth}
        \centering
        \includegraphics[width=\textwidth]{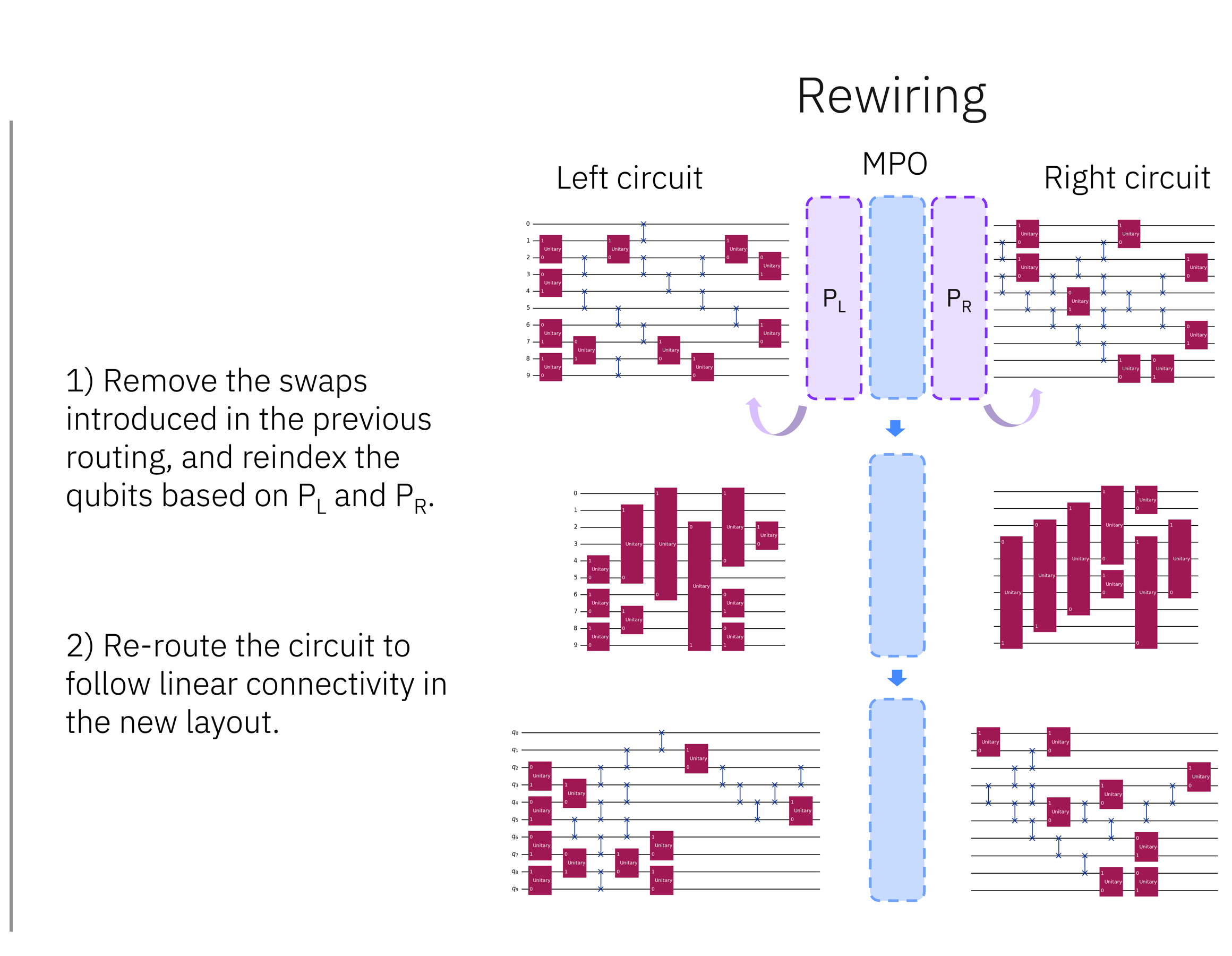}
        \caption{Rewiring}
        \label{fig:rewiring}
    \end{subfigure}
    \caption{The three stages of the iterative contraction method. (a)~The transpiled circuit is split at the temporal midpoint into a left circuit $C_L$ and a right circuit $C_R$, with an identity MPO inserted between them. (b)~The greedy unswapping procedure: qubit pairs in the MPO are ranked by bond dimension, and swaps are applied from the left, right, or both sides. Swaps that reduce the bond dimension are accepted, yielding the decomposition $M = P_L \tilde{M} P_R$. (c)~Rewiring: the extracted permutations $P_L$ and $P_R$ are absorbed into the remaining circuits by removing existing transpilation SWAPs, reindexing qubits, and re-transpiling to linear connectivity.}
    \label{fig:method_steps}
\end{figure*}

We then split the transpiled circuit into two halves, a left circuit $C_L$ and a right circuit $C_R$, at the temporal midpoint, and insert an empty (identity) Matrix Product Operator (MPO) between them. The full circuit is thus represented as $C_L \cdot M \cdot C_R$, where $M$ is initialized as the identity MPO. The left and right circuits are oriented so that their innermost layers (closest to the midpoint) are adjacent to the MPO, and contraction proceeds inward from both sides. This setup is illustrated in Fig.~\ref{fig:splitting}.

\subsection{Iterative Contraction}
\label{sec:contraction}

The core of the method is an iterative loop that progressively absorbs the left and right circuits into the central MPO. Each iteration consists of up to three stages: absorption, unswapping, and rewiring. The process repeats until all circuit layers have been absorbed. The full procedure is summarized in Algorithm~\ref{alg:contraction}.

\begin{figure}[t]
\refstepcounter{algorithm}
\label{alg:contraction}
\hrule\vspace{4pt}
\noindent\textbf{Algorithm \thealgorithm}\quad MPO Iterative Cancellation with Unswapping
\vspace{4pt}\hrule\vspace{4pt}
\begin{algorithmic}[1]
\Require Circuit $C$ on $N$ qubits, SVD cutoff $\epsilon$, MPO size threshold $\tau$
\Ensure MPS representation of $C |0\rangle^{\otimes N}$
\State Transpile $C$ to linear connectivity
\State Split $C$ at midpoint into left $C_L$ and right $C_R$
\State Initialize MPO $M \gets I^{\otimes N}$ (identity, bond dimension 1)
\While{$C_L$ or $C_R$ have remaining layers}
    \Comment{Absorption}
    \While{$\text{size}(M) < \tau$ \textbf{and} layers remain}
        \State Absorb innermost layer of $C_L$ into $M$ from left
        \State Absorb innermost layer of $C_R$ into $M$ from right
        \State Compress $M$ to canonical form via SVD with $\epsilon$
    \EndWhile
    \Comment{Unswapping}
    \State $P_L \gets I$, \quad $P_R \gets I$
    \State $A \gets \{0, 1, \ldots, N{-}2\}$ \Comment{available bond indices}
    \While{$A$ is not empty}
        \State $i \gets \arg\max_{j \in A} \text{bond\_dim}(M, j)$ \Comment{largest bond}
        \For{$\text{side} \in \{\text{left},\, \text{right},\, \text{both}\}$}
            \State $M' \gets \text{ApplySwap}(M, i, i{+}1, \text{side})$
            \State Compress $M'$ via SVD with cutoff $\epsilon$
        \EndFor
        \If{any variant reduced bond dim. at $(i, i{+}1)$}
            \State $M \gets$ best variant
            \State Update $P_L$ and/or $P_R$ with the applied swap
            \State $A \gets A \setminus \{i\}$ \Comment{mark current bond unavailable}
            \State $A \gets A \cup \text{neighbors}(i)$ \Comment{re-enable adj. bonds}
        \Else
            \State $A \gets A \setminus \{i\}$ \Comment{mark current bond unavailable}
        \EndIf
    \EndWhile
    \Comment{Rewiring}
    \State Remove transpilation SWAPs from $C_L$ and $C_R$
    \State Reindex qubits of $C_L$ according to $P_L$
    \State Reindex qubits of $C_R$ according to $P_R$
    \State Re-transpile $C_L$ and $C_R$ to linear connectivity
\EndWhile
\State Apply $M$ to $|0\rangle^{\otimes N}$ to obtain MPS $|\psi\rangle$
\State \Return $|\psi\rangle$
\end{algorithmic}
\vspace{4pt}\hrule
\end{figure}

\subsubsection{Absorption}

In the absorption stage, we contract the innermost layers of $C_L$ and $C_R$ into the MPO. Each two-qubit gate adjacent to the MPO is absorbed by contracting the gate tensor with the corresponding MPO site tensors, followed by a singular value decomposition (SVD) to restore the MPO to canonical form. A small singular value cutoff $\epsilon$ is applied during the SVD to discard negligible singular values and maintain numerical stability; this is the only source of approximation in the method.

Because the circuit has an underlying mirrored structure, many of the gates absorbed from opposite sides partially cancel, keeping the MPO bond dimension moderate during the early stages of absorption. However, as the swap-based obfuscation layers are encountered, the bond dimension begins to grow. When the total number of tensor elements in the MPO exceeds a predefined threshold $\tau$, we move to the unswapping stage.

\subsubsection{Unswapping}

The unswapping procedure is a greedy heuristic designed to extract hidden permutation structure from the MPO, thereby reducing its bond dimension. The key observation is that the swap obfuscation in the original circuit effectively permutes qubit lines within the MPO; if these permutations can be identified and factored out, the remaining MPO has a much smaller bond dimension.

The procedure maintains a set of \emph{available} qubit pairs (bonds), initialized to all $N-1$ nearest-neighbor pairs in the MPO. At each step, the available pairs are sorted by their current bond dimension, and the pair $(i, i+1)$ with the largest bond is selected. For this pair, three candidate swaps are evaluated: applying a SWAP from the left side of the MPO, from the right side, or from both sides simultaneously. Each candidate is followed by SVD compression. If the best candidate reduces the bond dimension, it is accepted: the MPO is updated, the swap is recorded in the corresponding permutation ($P_L$, $P_R$, or both), and the pair $(i, i+1)$ is marked as unavailable. Its adjacent pairs ($(i{-}1, i)$ and $(i{+}1, i{+}2)$, if they exist) are then marked as available, since the accepted swap changes the local structure of the MPO and may make neighboring bonds amenable to further reduction. If none of the three candidates reduces the bond dimension, the pair is simply marked as unavailable. The procedure repeats from the sorting step until no available pairs remain.

After unswapping, the original MPO is effectively decomposed as
\begin{equation}
    M = P_L \, \tilde{M} \, P_R,
\end{equation}
where $P_L$ and $P_R$ are the accumulated left and right permutations (each a product of SWAP gates), and $\tilde{M}$ is a reduced MPO with lower bond dimension. This decomposition is illustrated in Fig.~\ref{fig:unswapping}.

\subsubsection{Rewiring}

The permutations $P_L$ and $P_R$ extracted during unswapping must be absorbed into the remaining left and right circuits so that subsequent absorption steps operate on a consistent qubit ordering. This is accomplished through a rewiring procedure:
\begin{enumerate}
    \item \textbf{Remove existing SWAPs.} Strip all SWAP gates from $C_L$ and $C_R$ that were introduced during the initial transpilation to linear connectivity. This recovers the original qubit labeling of the remaining gates.
    \item \textbf{Reindex qubits.} Apply the permutations $P_L$ and $P_R$ to relabel the qubits in $C_L$ and $C_R$ respectively, so that the circuit qubit ordering matches the compressed MPO $\tilde{M}$.
    \item \textbf{Re-transpile.} Transpile the relabeled circuits back to linear connectivity with the new qubit layout. This reintroduces SWAP gates as needed but with a layout that is now aligned with the reduced MPO.
\end{enumerate}

After rewiring, the left and right circuits are again adjacent to the MPO with a consistent qubit ordering, and the absorption stage can resume. The rewiring procedure is illustrated in Fig.~\ref{fig:rewiring}.

\subsection{Sampling}
\label{sec:sampling}

Once the full circuit has been contracted into the MPO $M$, we obtain the output state by applying $M$ to the all-zero input state $|0\rangle^{\otimes N}$, yielding a Matrix Product State (MPS) representation of the output. Sampling from this MPS is efficient and exact: we sweep through the sites from left to right, sampling each qubit conditioned on the previously sampled qubits by contracting the conditional probability from the MPS tensors.

\section{Results}
\label{sec:results}

We apply the method described above to the \texttt{peaked\_circuit\_P9\_Hqap\_56x1917} circuit from Ref.~\cite{bluequbit_peaked}, a 56-qubit circuit with 1,917 two-qubit RZZ gates. This is the largest circuit for which the authors claim quantum advantage. The contraction was performed on a single Nvidia A100 GPU (80~GB) and completed in 4,059 seconds (approximately 1 hour and 10 minutes). For comparison, the same circuit was executed on Quantinuum's H2 trapped-ion processor in under 2 hours~\cite{bluequbit_peaked}, making our classical simulation roughly twice as fast as the quantum execution.

\subsection{Contraction Dynamics}
\label{sec:dynamics}

\begin{figure*}[t]
    \centering
    \begin{subfigure}[t]{\textwidth}
        \centering
        \includegraphics[width=\textwidth]{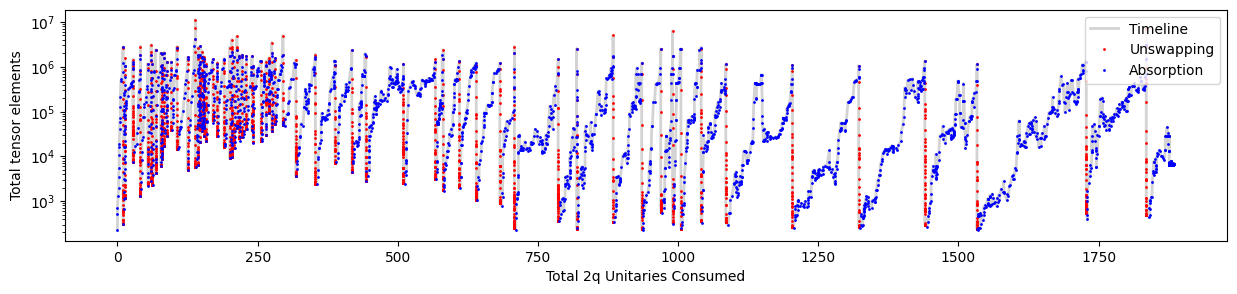}
        \caption{}
        \label{fig:elems_vs_unitaries}
    \end{subfigure}
    \vspace{0.5em}
    \begin{subfigure}[t]{\textwidth}
        \centering
        \includegraphics[width=\textwidth]{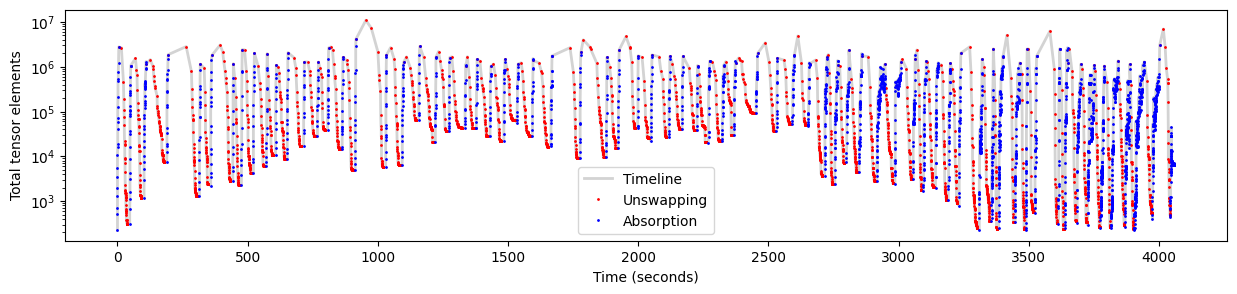}
        \caption{}
        \label{fig:elems_vs_time}
    \end{subfigure}
    \caption{Total number of tensor elements in the MPO during contraction. Blue points correspond to the absorption stage and red points to the unswapping stage. (a)~MPO size as a function of two-qubit unitaries consumed from the circuit. Three regimes are visible: an initial phase (0--300 unitaries) with rapid absorption--unswapping cycling; a transition phase (300--700) where unswapping becomes progressively more effective; and a final phase (700+) where unswapping reduces the MPO almost completely, allowing long absorption runs. (b)~The same quantity plotted against wall-clock time. The full contraction completes in 4,059 seconds on a single Nvidia A100 GPU. The dense cluster of iterations in the first ${\sim}1{,}000$ seconds reflects the initial phase where frequent unswapping calls dominate the runtime.}
    \label{fig:contraction_dynamics}
\end{figure*}

Fig.~\ref{fig:elems_vs_unitaries} shows the total number of tensor elements in the MPO as a function of the number of two-qubit unitaries absorbed from the circuit. The absorption (blue) and unswapping (red) stages produce a characteristic sawtooth pattern: the MPO grows during absorption as new gates are contracted in, then shrinks during unswapping as permutation structure is extracted.

The upper envelope of the MPO size is roughly determined by the unswapping threshold $\tau$: once the total number of elements exceeds $\tau$, absorption pauses and unswapping is triggered. This threshold is a tunable parameter, and the peaks in the sawtooth can overshoot it by some margin depending on the gates absorbed in the last absorption step. The key quantity that determines the effectiveness of the method is not the peak size, but rather how much the unswapping is able to reduce it. A large reduction means the absorption can run for many more layers before reaching the threshold again; in the worst case (for instance, on a random circuit with no mirror or permutation structure) unswapping would fail to reduce the MPO at all, and the algorithm would stall.

Three distinct regimes are visible in Fig.~\ref{fig:elems_vs_unitaries}. In the first phase (roughly 0--300 unitaries consumed), the absorption--unswapping cycle is rapid: the MPO reaches the threshold quickly, and unswapping only manages to reduce the size to around $10^4$ elements before the threshold is hit again. This results in frequent unswapping calls, which constitute overhead without proportional progress through the circuit. In the second phase (roughly 300--700 unitaries), a transition occurs: the absorption runs grow progressively longer, and the post-unswapping MPO size decreases steadily. In the third phase (beyond approximately 700 unitaries), the unswapping reduces the MPO almost completely at each iteration, allowing long absorption runs that consume many layers before the threshold is reached again.

A possible interpretation of this behavior is that the native obfuscation (the permutation structure introduced by the variational training and masking, rather than by explicit swap gates) is concentrated in the central region of the circuit, close to the midpoint of the $UU^\dagger$ identity block. During the early phase of contraction, this native permutation is only partially absorbed into the MPO, so the unswapping heuristic cannot fully invert it: the permutation may not decompose cleanly into nearest-neighbor swaps, for example if it was prepared in a way that it is spread across multi-qubit circuit blocks that were prepared variationally to encode a permutation implicitly. As more of this region is absorbed and the native obfuscation is fully captured by the MPO, the remaining permutation content comes predominantly from the explicit swap gates introduced by our own transpilation to linear connectivity. These artificial permutations are composed entirely of nearest-neighbor swaps by construction, and the unswapping procedure can cancel them efficiently.

Another factor that may contribute to the slow initial phase is that the splitting point may not coincide with the exact center of the underlying mirror circuit. Since we split at the temporal midpoint of the transpiled circuit, and the transpilation can shift the relative positions of gates, the left and right halves may not be perfectly symmetric at the start. This would produce a similar effect: the MPO can only be partially reduced because the two sides do not fully cancel. The transition phase may then correspond in part to the absorption process inadvertently correcting this misalignment, as the routing of layers to the linear topology can cause unitaries to be absorbed at different rates from the two sides, gradually shifting the effective splitting point toward the true mirror center.

Fig.~\ref{fig:elems_vs_time} shows the same quantity plotted against wall-clock time, confirming that the total contraction completes in approximately 4,059 seconds. The dense cluster of iterations in the first ${\sim}1{,}000$ seconds corresponds to the first phase described above, where frequent unswapping calls dominate the runtime despite limited progress through the circuit. The later phases proceed more efficiently, with absorption consuming the majority of the remaining layers.

\subsection{Peak Recovery}
\label{sec:peak_recovery}

\begin{figure}[t]
    \centering
    \includegraphics[width=\columnwidth]{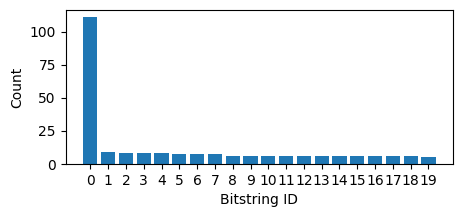}
    \caption{Frequency of the top 20 most-sampled bitstrings from 1,000 samples drawn from the contracted MPS. The peak bitstring (ID~0) appears approximately 110 times (${\sim}11\%$), consistent with the designed peak weight of ${\sim}10\%$. The sharp separation from the remaining bitstrings confirms successful recovery of the peak.}
    \label{fig:histogram}
\end{figure}

After contracting the full circuit, we apply the resulting MPO to the all-zero state $|0\rangle^{\otimes 56}$ to obtain an MPS and draw 1,000 samples. Figure~\ref{fig:histogram} shows the frequency of the top 20 most-sampled bitstrings. The peak bitstring (ID~0) appears approximately 110 times out of 1,000 samples, corresponding to a measured peak weight of roughly 11\%. This is consistent with the designed peak weight of approximately 10\% reported in Ref.~\cite{bluequbit_peaked}, and confirms that the contraction has preserved the circuit's output distribution to high accuracy. The remaining bitstrings each appear fewer than 20 times, with no secondary structure visible in their frequencies. This sharp separation between the peak and the background is the defining signature of the peaked circuit and demonstrates that our method successfully recovers it.

The only source of approximation in our method is the small singular value cutoff $\epsilon$ applied during SVD compression. The close agreement between our measured peak weight (${\sim}11\%$) and the designed value (${\sim}10\%$) provides empirical evidence that this cutoff introduces negligible error for the circuit sizes considered here. In principle, the cutoff can be made arbitrarily small at the cost of increased MPO bond dimension; in practice, we find that a modest cutoff is sufficient to maintain near-exact fidelity throughout the contraction.

\section{Discussion}
\label{sec:discussion}

The effectiveness of our approach rests on a single structural feature of the peaked circuits in Ref.~\cite{bluequbit_peaked}: the $UU^\dagger$ mirror construction. Regardless of the obfuscation applied, the two halves of the circuit remain approximate inverses of each other, and contracting them simultaneously into a central MPO allows this cancellation to take place implicitly. The obfuscation techniques are designed to prevent classical methods from recognizing and exploiting this structure. However, our results show that the swap-based permutation obfuscation, which is the primary barrier to MPO-based contraction, can be effectively undone through the greedy unswapping procedure without requiring any knowledge of the original permutation.

This points to a fundamental tension in the peaked circuit construction. The mirror structure is what makes the circuit peaked: the $UU^\dagger$ block ensures that the output distribution is concentrated on a known bitstring. But this same structure is also what makes the circuit classically simulable, because it guarantees that the two halves cancel when contracted from the midpoint. The obfuscation must therefore hide the mirror structure well enough to defeat classical simulation while preserving it well enough to maintain the peak. Our results suggest that, for the constructions considered here, this balance has not been achieved.

\subsection{Scope and Limitations}

It is important to distinguish between the general peaked circuit problem and the specific HQAP construction of Ref.~\cite{bluequbit_peaked}. The authors prove that deciding whether a circuit is peaked is QCMA-complete in the worst case, and conjecture that finding the peak bitstring is hard even for quantum algorithms given only the circuit description. Our work does not contradict these complexity-theoretic results. Rather, it shows that the specific construction used to generate the circuits in Ref.~\cite{bluequbit_peaked} which relies on a mirror circuit with obfuscated permutations does not achieve the hardness that the general problem admits.

Our method is also specifically tailored to circuits with this mirror structure. It would not be effective against peaked circuits constructed through fundamentally different means, nor against generic quantum circuits. As noted in the results, applying the method to a random circuit (without underlying permutation structure) would cause the unswapping to fail, and the MPO bond dimension would grow uncontrollably. The method succeeds here precisely because the peaked circuit construction leaves exploitable structure in the form of permutations that can be extracted.

One might consider modifications to the HQAP construction that could increase the classical cost. While such modifications could increase the runtime of our method or require adaptations to the unswapping heuristic, we note that they do not address the underlying vulnerability: as long as the circuit is built from a mirror structure, the two halves will partially cancel during contraction, and extensions of the present approach can exploit this. It is unclear whether these improvements can achieve a superpolynomial separation. However, they may increase the constant prefactor of the classical simulation time enough that a direct execution in a quantum device achieves better wall-clock time, resulting in a practical instance of quantum advantage.

\subsection{Towards better verifyable quantum advantage}

More broadly, our results suggest that the vulnerability of the HQAP construction lies not in the specific obfuscation being too weak, but in the peaking mechanism being too structured. In mirror-based constructions, the peak is \emph{designed in} by ensuring that $UU^\dagger$ acts as the identity on the input state, so the output is predetermined by construction. Any circuit built this way carries structural information that a sufficiently clever classical algorithm can exploit, regardless of how well the structure is obfuscated. 

A potentially more robust approach would be to construct circuits where the peak \emph{emerges} from the computational hardness of an encoded problem, rather than from a designed structural shortcut. In such a construction, the classical hardness of finding the peak would be inherited from the hardness of the underlying problem itself, not from the difficulty of reverse-engineering the circuit. 

Shor's algorithm~\cite{shor} provides a clean example of this principle: its output distribution is peaked on bitstrings encoding the period, and finding the period is believed to be classically hard. However, Shor's algorithm requires fault-tolerant quantum hardware and is far from near-term implementation. A more promising direction for near-term verifiable quantum advantage may lie in circuits that encode classically hard optimization or combinatorial problems whose solutions can be efficiently verified, such as the problems described in ~\cite{intractable_decathlon}. If such circuits can be designed to produce sufficiently peaked output distributions on solution bitstrings, they would combine verifiability (the solution can be checked) with hardness that does not depend on obfuscation. An example of this paradigm is the Hidden Code Sampling (HCS) proposal of Deshpande et al.~\cite{deshpande_hcs}, which constructs peaked distributions from quantum error correction structure rather than mirror circuits, with classical hardness tied to computing weight enumerators of random codes. 

\subsection{Broader Context}

This work adds to the pattern of quantum advantage claims being met by classical algorithmic improvements that narrow or close the gap. The peaked circuit framework of Ref.~\cite{bluequbit_peaked} was explicitly designed with this iterative dynamic in mind, framing the circuits as a public challenge analogous to the RSA Factoring Challenge. Our results have been submitted to the Quantum Advantage Tracker~\cite{qa_tracker}, where they appear alongside the quantum execution results and the original classical extrapolations under the classically verifiable problems pathway. In this spirit, our result should be understood as a response to that challenge: the specific instances presented do not withstand classical simulation, but the broader question of whether peaked circuits can be constructed to resist all classical attacks remains open.

\subsection{Code Availability}

The source code for the MPO contraction and unswapping method is available at \url{https://github.com/d-kremer/peaked-circuit-simulation}.

\bibliography{references}

\clearpage
\onecolumngrid
\appendix

\section{Implementation Details and Algorithm Variants}
\label{app:variants}

The algorithm presented in the main text (Algorithm~\ref{alg:contraction}) describes the conceptual structure of the method. The implementation used to produce the results in this paper includes several refinements that improve practical performance. We describe these here for reproducibility.

\subsection{Parallel Swap Selection}
\label{app:parallel}

The unswapping procedure described in the main text evaluates swap candidates sequentially: the bond with the largest dimension is tested, accepted or rejected, and the process restarts. In practice, we use a parallelized variant that evaluates multiple swap candidates simultaneously.

The key observation is that swaps acting on non-overlapping qubit pairs are independent and can be evaluated in parallel. We partition the bonds into two sets by parity (even pairs $(0,1), (2,3), \ldots$ and odd pairs $(1,2), (3,4), \ldots$) so that all pairs within a set are non-overlapping. For each combination of side (left, right, both) and parity (even, odd), we apply all candidate swaps simultaneously and retain only those that reduce the local bond dimension. The six combinations are cycled through in sequence: both-even, both-odd, left-even, left-odd, right-even, right-odd. The outer loop terminates when a full cycle produces no further bond reductions.

This parallelized variant is significantly faster in practice, since each iteration tests $\lfloor N/2 \rfloor$ swaps at once rather than one at a time. Because the even/odd partition ensures that the swaps within each batch act on disjoint qubit pairs, the bond dimension reduction from each individual swap is independent of the others within the same batch.

\subsection{Non-Strict Acceptance Criterion}
\label{app:equal}

The default acceptance criterion for unswapping accepts a swap only if it strictly reduces the bond dimension at the target bond. We also implement a relaxed variant that accepts swaps that leave the bond dimension unchanged (while still rejecting those that increase it). This ``equal'' mode can help the algorithm escape local minima where a permutation requires multiple swaps to resolve but no single swap produces an immediate improvement.

\subsection{Adaptive Side Selection}
\label{app:adaptive}

In the main text, the absorption stage alternates between absorbing from the left and right circuits. The implementation supports an adaptive variant in which, at each absorption step, we tentatively absorb the next layer from both $C_L$ and $C_R$ independently and select the side that results in a smaller total MPO size. This greedy selection helps maintain a balanced contraction and can avoid situations where one side grows the MPO disproportionately. An alternative fixed-schedule mode absorbs from each side at a configurable frequency, which can be useful when the adaptive overhead (two trial absorptions per step) is not justified.

\subsection{Hyperparameters}
\label{app:hyperparameters}

The method has the following tunable hyperparameters:

\begin{itemize}
    \item \textbf{SVD cutoff} ($\epsilon$): threshold for discarding singular values during MPO compression. Controls the approximation quality. We use $\epsilon = 2 \times10^{-3}$ for the results in this paper.
    \item \textbf{Maximum bond dimension} ($\chi_{\max}$): hard cap on the bond dimension after SVD compression. We use $\chi_{\max} = 8192$.
    \item \textbf{Unswapping threshold} ($\tau$): total number of MPO tensor elements at which absorption pauses and unswapping is triggered. We use $\tau = 10^6$.
    \item \textbf{Maximum unswapping iterations}: cap on the number of outer-loop iterations in the unswapping procedure. We use 20 iterations.
    \item \textbf{Acceptance mode}: strict (accept only dimension-reducing swaps) or relaxed (also accept dimension-preserving swaps). We use strict mode.
    \item \textbf{Side selection}: adaptive (greedy selection based on MPO size) or fixed-frequency alternation. We use adaptive selection.
\end{itemize}

The results reported in the main text were obtained with the hyperparameters listed above. We did not perform an extensive hyperparameter search; the values were chosen based on preliminary experiments and may not be optimal. A systematic study of the sensitivity of the method to these hyperparameters is left for future work.

\end{document}